\documentclass[twocolumn]{jpsj2} 
\def\nle{\ \raise.3ex\hbox{$<$}\kern-0.8em\lower.7ex\hbox{$\sim$}\ }
\def\nge{\ \raise.3ex\hbox{$>$}\kern-0.8em\lower.7ex\hbox{$\sim$}\ }

\newcommand{\mattt}[1]{\left(\begin{array}{@{\,}ccc@{\,}} #1 \end{array}\right)} 
\newcommand{\matt}[1]{\left(\begin{array}{@{\,}cc@{\,}} #1 \end{array}\right)} 
\newcommand{\lefbrace}[1]{\left\{\begin{array}{@{\,}l} #1 \end{array}\right.}

\title{Spin-Wave Theory of the Multiple-Spin Exchange Model on a
Triangular Lattice in a Magnetic Field : 3-Sublattice Structures}

\author{\textsc{Chitoshi Yasuda}\thanks{E-mail address:
cyasuda@phys.aoyama.ac.jp}, \textsc{Daisuke Kinouchi}
and \textsc{Kenn Kubo}}

\inst{Department of Physics and Mathematics, Aoyama Gakuin University,
Sagamihara 229-8558, Japan}

\abst{
We study the spin wave in the $S=1/2$ multiple-spin exchange model on a
triangular lattice in a magnetic field within the linear spin-wave
theory. We take only two-, three- and four-spin exchange  
interactions into account and restrict ourselves to the region 
where a coplanar three-sublattice state is the mean-field ground state. 
We found that the Y-shape ground state survives quantum fluctuations and
the phase transition to a phase with a 6-sublattice structure occurs
with softening of the spin wave. 
We estimated the quantum corrections to the ground state sublattice
magnetizations due to zero-point spin-wave fluctuations.}

\kword{multiple-spin exchange model, three-sublattice structure,
spin-wave approximation, softening, sublattice magnetization, 
quantum phase transition, solid helium, frustration}

\begin{document}
\maketitle

\section{Introduction}

The multiple-spin exchange (MSE) model on the triangular lattice is one
of the two dimensional frustrated quantum spin systems which are 
presently at the focus of strong interest. The model is believed to
describe nuclear magnetism of the solid $^3$He layers adsorbed on
graphite~\cite{Franco,Godfrin,Siqueira}, whose peculiar temperature
dependence of the specific heat with a double-peaked structure is quite
intriguing~\cite{Ishida}. Several studies were devoted for theoretical
understanding of the experimental results in terms of the MSE
model~\cite{Kubo,Kubo2,Momoi,Misguich,Misguich2,Misguich3,Momoi2,Roger2,Kubo3}.
The mean-field theory was applied to the MSE model with two-, three-
and four-spin cyclic exchange interactions, which lead to various kinds of 
the ground state corresponding to different values of
parameters~\cite{Kubo,Kubo3}.
Appearance of various mean-field ground states reflects 
weak stability of the mean-field ground states due to strong frustration. 
Therefore quantum fluctuations should have strong influence on the
ground state of the system. Quantum mechanical ground state was studied
by the numerical diagonalization of finite clusters and a spin liquid
ground state with a finite spin gap was predicted for the parameters
which are thought to be adequate to the $^3$He
layer~\cite{Misguich2,Misguich3}. However recent measurement of the
susceptibility down to $\sim $10 $\mu$K did not show any sign which
suggests the existence of a spin gap~\cite{Masutomi}. At present, the
ground state of both the two-dimensional solid $^3$He on graphite and
the MSE model on the triangular lattice are, therefore, hardly
understood. The MSE interactions might be relevant as well to the spin
liquid state of $\kappa$-(ET)$_2$Cu$_2$(CN)$_3$ where antiferromagnetic
interaction is thought to be dominant~\cite{Kurosaki}.   
Therefore it is meaningful and desired to study the properties of the
MSE model on the triangular lattice into detail.
    
\begin{figure}[tb]
  \centerline{\resizebox{0.3\textwidth}{!}{\includegraphics{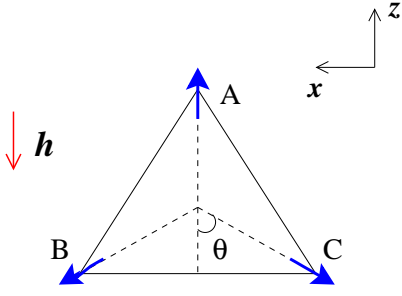}}}
  \caption{Y-shape state on a triangular lattice in a magnetic field
 {\boldmath $h$}.}
  \label{def0}
\end{figure}

In the present work, we study the spin wave in the $S=1/2$ MSE model
with two-, three- and four-spin exchange interactions on the triangular
lattice in the magnetic field. The system is described by the
Hamiltonian 
\begin{equation}
  {\cal H} = J \sum_{<i,j>} \mib{\sigma}_i \cdot \mib{\sigma}_j
           + K \sum_{\rm p} h_{\rm p} + h \sum_i \sigma_i^z \ ,
\label{ham}
\end{equation}
where $\mib{\sigma}_i$ is the Pauli matrix on the site $i$ and the
summations $\sum_{<i,j>}$ and $\sum_{\rm p}$ run over all pairs and
minimum diamonds, respectively. Since the three-spin exchange
interaction is reduced to the conventional two-spin one, the first term
of the Hamiltonian~(\ref{ham}) describes the sum of two- and three-spin
exchange interactions. The Hamiltonian $h_{\rm p}$ is the four-spin
exchange interaction for a minimum diamond p. For a diamond of spins 1
$\sim$ 4 with diagonal bonds (1,3) and (2,4), it reads
\begin{eqnarray}
  h_{\rm p} &=& \sum_{1 \le i < j \le 4} \mib{\sigma}_i \cdot 
            \mib{\sigma}_j
           + (\mib{\sigma}_1 \cdot \mib{\sigma}_2)
             (\mib{\sigma}_3 \cdot \mib{\sigma}_4) \nonumber \\ 
         &+& (\mib{\sigma}_1 \cdot \mib{\sigma}_4) 
             (\mib{\sigma}_2 \cdot \mib{\sigma}_3) 
           - (\mib{\sigma}_1 \cdot \mib{\sigma}_3) 
             (\mib{\sigma}_2 \cdot \mib{\sigma}_4)  \ .
\end{eqnarray}
The coupling parameters are written in terms of conventional exchange
constants as $K=-J_4 /4$ and $J= J_3 - J_2 /2$. Since $J_n$ are known to
be negative in solid $^3$He~\cite{Thouless}, we take the value of $K$
positive in the following. The third term in Eq.~(\ref{ham}) is the
Zeeman term with a magnetic field $\mib{h}$ ($h = |\mib{h}|$) applied
anti-parallel to the $z$-direction as shown in Fig.~\ref{def0}.

Numerous mean-field ground-state phases were found for the
Hamiltonian~(\ref{ham}) by assuming up to 144 sublattices~\cite{Kubo3}.
The phase diagram is parametrized by $J/K$ and $h/K$. When $h=0$, the
ground state varies according to $J/K$ as follows : I) the ferromagnetic
phase for $J/K <-8.61$. II) the intermediate phase for 
$-8.61 < J/K < -2.26$ where the ground state spin structure varies with
the change of $J/K$ and ten small phases with equal to or more than 12
sublattices were identified. These small phases might be artifacts of
assumptions of finite number of sublattices. The true ground state in
this parameter region is still controversial even at the mean-field
level. III) the tetrahedral phase for $-2.26 < J/K <8.22$. The ground
state has a 4-sublattice spin structure with zero magnetization, where
spin vectors on four sublattices point four vertices of a tetrahedron if
their bottoms are put at its center. IV) a phase with six sublattices
for $8.22 < J/K < 10$. The ground state has non-coplaner spin structure
with uniform vector chiral order and staggered scalar chiral order. 
Finally, V) the 120$^{\circ}$ phase for $J > 10K$. The spins on three
sublattices make the angle $2\pi/3$ to each other. 

The mean-field phase diagram in the magnetic field was also
studied\cite{Kubo,Momoi2,Kubo3}. One of the interesting behaviors is the
appearance of a magnetization plateau with 1/2 of the full polarization
in the parameter region adequate to $^3$He
layers~\cite{Kubo,Misguich2,Momoi2}. This plateau is realized by the
4-sublattice uuud spin structure. Other numerous ground-state phases in
the magnetic field have been found, but we introduce only the phases
related to the present work. The 120$^{\circ}$ ground state for $J >10K$
is modified by the magnetic field. This state has a 3-sublattice
coplanar spin configuration with the spins on one (say A) sublattice
antiparallel to the applied magnetic field. The spins on other
sublattices (B and C) tilt toward the oblique directions as shown in
Fig.~\ref{def0}. In the present paper, we call it Y-shape
state~\cite{y-shape}. This result shows that the four-spin exchange
interaction lifts the non-trivial degeneracy of the mean-field 
ground-state of the antiferromagnetic Heisenberg (AFH) model in the
magnetic field~\cite{Kawamura}. A 6-sublattice phase exists for 
$8\nle J/K \nle 10$ adjacent to the Y-shape state for weak magnetic
field. Increase in the magnetic field induces a 12-sublattice
structure. Further increase in the magnetic field leads to a
magnetization plateau with 1/3 of full polarization due to the
3-sublattice uud state with two sublattices with up spins and the other
with down spins. The uud state is stabilized by the four-spin interaction
and occupies a large region in the phase diagram. We note that the phase
boundary between the Y-shape and the 6-sublattice phases was not
determined by the mean-field theory.

Spin waves in this model were studied previously in the absence of the
magnetic field~\cite{Momoi,Kubo2, Momoi2}. The spin wave in the
tetrahedral phase was studied and the quantum corrections to the
sublattice magnetization were estimated~\cite{Momoi}. The analysis for
$J=0$ showed the stability of the tetrahedral state against the
zero-point fluctuations of the spin wave. The spin wave in the
120$^\circ$ phase was studied and it was shown that there are three
gapless branches for small $k$ and the spin wave softens at
$J=10K$~\cite{Kubo2}. The softening corresponds to the phase transition
to the 6-sublattice phase. 

In the present work, we investigate the spin wave in the Y-shape phase 
and discuss the phase transition under the magnetic field. 
Although the Y-shape phase does not correspond to the model for the
solid $^3$He layer~\cite{Bernu}, coplay of the geometrical frustration
and the MSE interactions poses a problem of interest in itself. We
investigated the dispersions of the spin waves as well as the effects of
the zero-point fluctuations of the spin wave on the ground-state
energies. As a result, we found that though the Y-shape ground state
survives quantum fluctuations for $J/K \nge 12$ and small $h/K$,
softening of the spin wave leads to the phase transition to the
6-sublattice phase. 
     
The present paper is organized as follows. In \S~2, we present the
linear spin-wave theory for the Y-shape state in the magnetic
field. In \S~3, we simply summarize the magnetic properties in two
special cases, i.e., the AFH model in the magnetic field and the MSE
model at $h=0$. The dispersion of the spin wave and the phase diagram
obtained from the softening of the spin wave are shown in \S~4 and 5,
respectively. The quantum effects are discussed by studying quantum
corrections of the ground-state energy and the sublattice magnetizations
in \S~6. Finally, \S~7 is devoted to summary and discussion.

\section{Spin-Wave Theory for the Y-Shape State}  

Assuming the Y-shape state as the ground state, we can perform the
Holstein-Primakoff transformation~\cite{Holstein} of the
Hamiltonian~(\ref{ham}) by neglecting higher-order terms as 
\begin{equation}
  \lefbrace{
    \sigma_i^z \simeq 1 - 2 a_i^{\dagger} a_i \\
    \sigma_i^+ \simeq a_i \\
    \sigma_i^- \simeq a_i^{\dagger} 
  } \ ,
\label{HP_A}
\end{equation}
for $i \in$ A sublattice,
\begin{equation}
  \lefbrace{
    \sigma_j^z \simeq -\alpha (b_j^{\dagger} + b_j) 
                 -\beta (1-2b_j^{\dagger} b_j) \\
    \sigma_j^+ \simeq \frac{1}{2} \{\alpha (1-2b_j^{\dagger} b_j) 
             -(\beta+1) b_j^{\dagger}-(\beta-1) b_j \} \\
    \sigma_j^- \simeq \frac{1}{2} \{\alpha (1-2b_j^{\dagger} b_j) 
             -(\beta-1) b_j^{\dagger}-(\beta+1) b_j \}
  } \ ,
\label{HP_B}
\end{equation}
for $j \in$ B sublattice, and
\begin{equation}
  \lefbrace{
    \sigma_k^z \simeq \alpha (c_k^{\dagger} + c_k) 
                 -\beta (1-2c_k^{\dagger} c_k) \\
    \sigma_k^+ \simeq \frac{1}{2} \{-\alpha (1-2c_k^{\dagger} c_k) 
             -(\beta+1) c_k^{\dagger}-(\beta-1) c_k \} \\
    \sigma_k^- \simeq \frac{1}{2} \{-\alpha (1-2c_k^{\dagger} c_k) 
             -(\beta-1) c_k^{\dagger}-(\beta+1) c_k \}
  } \ ,
\end{equation}
for $k \in$ C sublattice,
where $a_i^{\dagger}$ ($a_i$), $b_j^{\dagger}$ ($b_j$) and
$c_k^{\dagger}$ ($c_k$) are the boson creation (annihilation)
operators, $\alpha=\sin{\theta}$ and $\beta=\cos{\theta}$ as shown in
Fig.~\ref{def0}.

First, we obtain the relation between $h$ and the angle $\theta$ from the
condition that the first-order terms of the transformed Hamiltonian
should vanish, i.e.,
\begin{equation}
  \cos{\theta} = \frac{1}{12K} \{ 5K+J-\sqrt{(J-K)^2-4Kh}\} \ ,
\label{kakudo}
\end{equation}
for $h \le h_{\rm c}$ where 
\begin{equation}
\label{eq:hc}
h_{\rm c}=-12K+3J. 
\end{equation}
The value of $\theta$ monotonously decreases with the increase of $h$
and the uud state ($\theta =0$) is realized for $h \ge h_{\rm c}$.
This relation can be also obtained from 
${\rm d} E^{\rm cl}/{\rm d} \theta = 0$ where the mean-field ground-state 
energy $E^{\rm cl}$ per the total number of sites $N$ is given by 
\begin{eqnarray}
\label{classical_E}
  && E^{\rm cl}/N = (2\beta^2-2\beta-1)J  \\
  && \hspace*{1cm} - (8\beta^3-10\beta^2+4\beta+1)K 
               + \frac{1}{3}(1-2\beta)h \ . \nonumber 
\end{eqnarray}

\begin{figure}[tb]
  \centerline{\resizebox{0.25\textwidth}{!}{\includegraphics{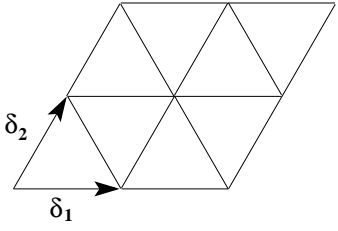}}}
  \caption{Unit vectors $\delta_1=(1,0)$ and $\delta_2=(0,1)$ on the
 triangular lattice.}
  \label{unit}
\end{figure}

After straightforward calculations the Hamiltonian is rewritten as
\begin{equation}
  {\cal H} = E^{\rm cl} 
           + \sum_{\mib{k}}' \mib{v_k}^{\dagger} {\cal D} \mib{v_k}
           + E_0^{\rm q} \ ,
\label{ham2}
\end{equation}
where $\mib{v_k}^{\dagger}=(a_{\mib{k}}^{\dagger},b_{\mib{k}}^{\dagger},
c_{\mib{k}}^{\dagger},a_{-\mib{k}},b_{-\mib{k}},c_{-\mib{k}})$ and
${\displaystyle \sum_{\mib{k}}' }$ in the second term denotes the
summation over a half of the reduced Brillouin zone of the 3-sublattice
structure. The matrix ${\cal D}$ reads
\begin{equation}
  {\cal D} = \matt{ M_{\rm diag} & M_{\rm off} \\
                    M_{\rm off} & M_{\rm diag} } \ ,
\end{equation}
where
\begin{equation}
  M_{\rm diag} = \mattt{ 
          A_{\mib{k}} & C \Gamma_{\mib{k}} & C \Gamma_{\mib{k}}^* \\
          C \Gamma_{\mib{k}}^* & B_{\mib{k}} & D \Gamma_{\mib{k}} \\
          C \Gamma_{\mib{k}} & D \Gamma_{\mib{k}}^* & B_{\mib{k}} } \ ,
\end{equation}
and
\begin{equation}
  M_{\rm off} = \mattt{ 
          E_{\mib{k}} & G \Gamma_{\mib{k}} & G \Gamma_{\mib{k}}^* \\
          G \Gamma_{\mib{k}}^* & F_{\mib{k}} & H \Gamma_{\mib{k}} \\
          G \Gamma_{\mib{k}} & H \Gamma_{\mib{k}}^* & F_{\mib{k}} } \ ,
\end{equation}
with
\begin{eqnarray}
 && A_{\mib{k}}=12 \{ -2K +(J+2K)\beta +4K\beta^3 \nonumber \\ 
 && \hspace*{1cm}  + K(1-\beta^2)\Delta_{\mib{k}} \} -2h \ , \nonumber \\
 && B_{\mib{k}}=6 \{ J+2K +(J-4K)\beta -2(J+5K)\beta^2 \nonumber \\
 && \hspace*{1cm} +12K\beta^3 +2K(1-\beta+2\beta^3)\Delta_{\mib{k}} \}
  +2h\beta \ , \nonumber \\
 && C=3(1-\beta)(J+4K+4K\beta^2) \ , \nonumber \\
 && D=6 \{ K(4\beta-1) +(J+5K)\beta^2 -8\beta^3K  \} \ , \nonumber \\
 && E_{\mib{k}}= -12(1-\beta^2)K\Delta_{\mib{k}} \ , \\
 && F_{\mib{k}}= 2\beta E_{\mib{k}} \ , \nonumber \\
 && G= -3(1+\beta)(J+4\beta^2K) \ , \nonumber \\
 && H= -6(1-\beta^2)(J+5K-8\beta K) \ . \nonumber
\end{eqnarray}
Here the wave-number dependence is incorporated by 
\begin{equation}
 \Gamma_{\mib{k}}=\frac{1}{3}({\rm e}^{{\rm i}k_1} 
      + {\rm e}^{-{\rm i}k_2} + {\rm e}^{{\rm i}(-k_1+k_2)} ) \ ,
\end{equation}
\begin{equation}
 \Delta_{\mib{k}}=\frac{1}{3} \{ \cos{(k_1+k_2)} +\cos{(2k_1-k_2)}
             +\cos{(-k_1+2k_2)} \} \ ,
\end{equation}
where $k_i = \mib{k}\cdot \mib{\delta}_i $¡Ê$\ i=1, 2$ ¡Ëis 
an element of the wave vector and the unit vector $\mib{\delta}_1 $ and
$\mib{\delta}_2 $ of the triangular lattice are chosen as shown in
Fig.~\ref{unit}. Note that $k_1^2 + k_2^2 -k_1 k_2 = 3k^2 /4$ holds.  
The third term in Eq.~(\ref{ham2})
\begin{equation}
  E_0^{\rm q}=-\sum_{\mib{k}}' (A_{\mib{k}}+2B_{\mib{k}}) \ 
\end{equation}
arises from the Bose commutation relation. 
The Hamiltonian~(\ref{ham2}) is diagonalized through the Bogoliubov
transformation and takes the following form:
\begin{equation}
 {\cal H} = E^{\rm cl}
          + \sum_{\mib{k}}' \mib{u_k}^{\dagger} {\cal D}_{\rm diag} \mib{u_k}
          + E_0^{\rm q} \ ,
\end{equation}
where
$\mib{u_k}^{\dagger}=(\alpha_{\mib{k}}^{\dagger},\beta_{\mib{k}}^{\dagger}, 
\gamma_{\mib{k}}^{\dagger},\alpha_{-\mib{k}},\beta_{-\mib{k}}, 
\gamma_{-\mib{k}})$  is a vector of the transformed Bose operators and 
${\cal D}_{\rm diag}$ is a diagonal matrix whose $(i, i)$ elements are 
$\omega_{\mib{k}}^{(1)}, \omega_{\mib{k}}^{(2)}, \omega_{\mib{k}}^{(3)}, 
\omega_{\mib{k}}^{(1)}, \omega_{\mib{k}}^{(2)}$ and
$\omega_{\mib{k}}^{(3)}$ for $i=1\sim 6$. The frequencies of the three
branches of the spin wave are given by $\omega_{\mib{k}}^{(1)}$,
$\omega_{\mib{k}}^{(2)}$ and $\omega_{\mib{k}}^{(3)}$.
We numerically performed the transformation according to the general
theory by Colpa~\cite{colpa}. 

\begin{figure}[td]
  \centerline{\resizebox{0.4\textwidth}{!}{\includegraphics{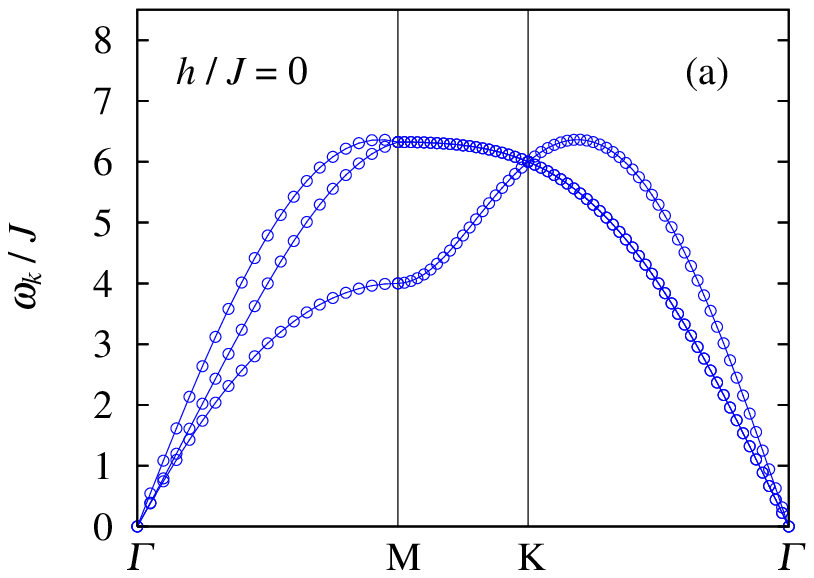}}}
  \centerline{\resizebox{0.4\textwidth}{!}{\includegraphics{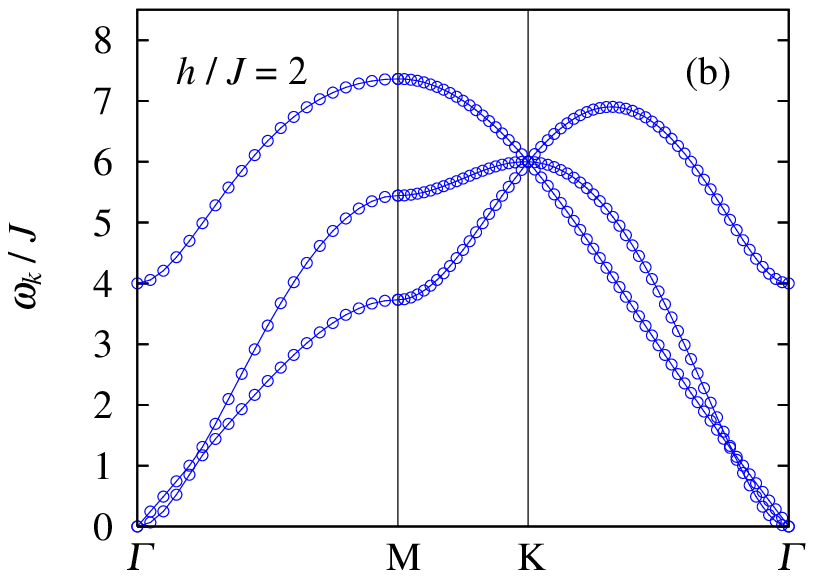}}}
  \caption{Three dispersions of the spin wave for the AFH model with (a)
 $h=0$ and (b) $h/J=2$. The marks $\Gamma$, M and K denote the wave
 numbers as defined in Fig.~\ref{zone} and the vertical axis the lines
 $\Gamma$-M-K-$\Gamma$.}
  \label{disp_K=0}
\end{figure}

\section{Spin-Wave Spectrum of the AFH Model in the Magnetic Field and
 the MSE Model at $h=0$}

In this section we reproduce the previously known results for simple cases 
in order for comparison with the results in the next section. 

\begin{figure}[t]
  \centerline{\resizebox{0.2\textwidth}{!}{\includegraphics{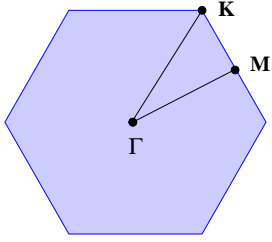}}}
  \caption{Brillouin zone of the 3-sublattice structure with reciprocal
 lattice vectors ($k_1$, $k_2$) = ($2\pi/3$, $-2\pi/3$) and ($2\pi/3$,
 $4\pi/3$). The marks $\Gamma$, M and K denote ($k_1$, $k_2$) = (0,0),
 ($2\pi/3$, $\pi/3$) and ($2\pi/3$, $2\pi/3$), respectively.}
  \label{zone}
\end{figure}

First we demonstrate the results for $K=0$, i.e., the AFH model.  
The dispersions of the spin waves assuming the Y-shape state as the
ground state for the AFH model with $h/J=0$ and 2 are shown in
Figs.~\ref{disp_K=0} (a) and (b), respectively. The hexagonal Brillouin
zone for the 3-sublattice structure is shown in Fig.~\ref{zone}. 
For $h=0$ the frequencies of all spin-wave branches vanish at
the $\Gamma$ point ($\mib{k}=0$) and have the linear dispersion 
for small $k$. In the magnetic field, one of the three branches is
lifted to $\omega_{\mib{k}}=2h$ at the $\Gamma$ point corresponding to 
the uniform precession of the spins about the magnetic field. 
There are two gapless branches rather than one expected from the SO(2)
symmetry. Furthermore one of them has a quadratic dispersion for small
$k$. This is caused by a non-trivial continuous degeneracy in the 
mean-field ground state in the magnetic field~\cite{Kawamura}.  
Any three-sublattice spin structure fulfilling the relation
$\mib{S}_{\rm A}+\mib{S}_{\rm B}+\mib{S}_{\rm C}=\mib{h}/3J$ 
is a ground state for $h \le 3J$, where $\mib{S}_{\rm A,B,C}$ is the
sublattice magnetization on each sublattice. 
Chubukov and Golosov showed that quantum fluctuations lift this
non-trivial degeneracy and select the coplanar spin structure, i.e., the
Y-shape state~\cite{Chubukov3}. This is an example of the so-called
order from disorder phenomena.   
  
\begin{figure}[td]
  \centerline{\resizebox{0.4\textwidth}{!}{\includegraphics{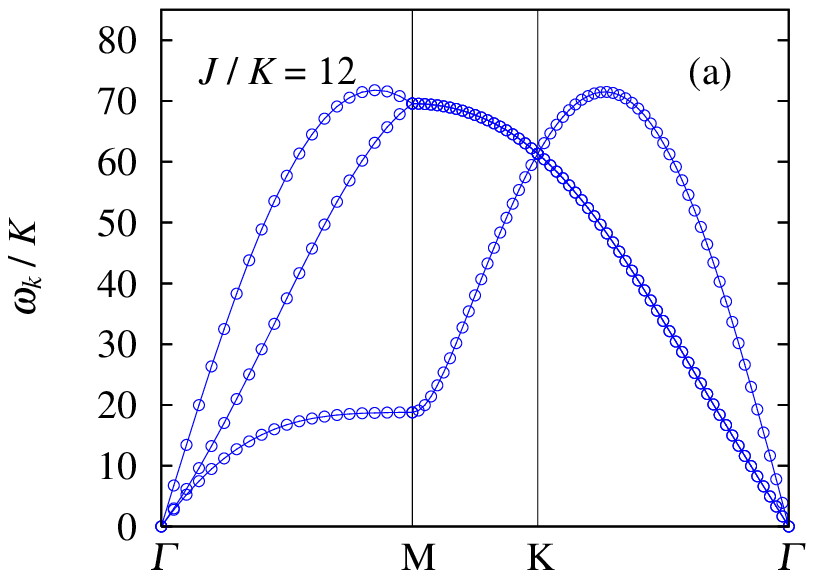}}}
  \centerline{\resizebox{0.4\textwidth}{!}{\includegraphics{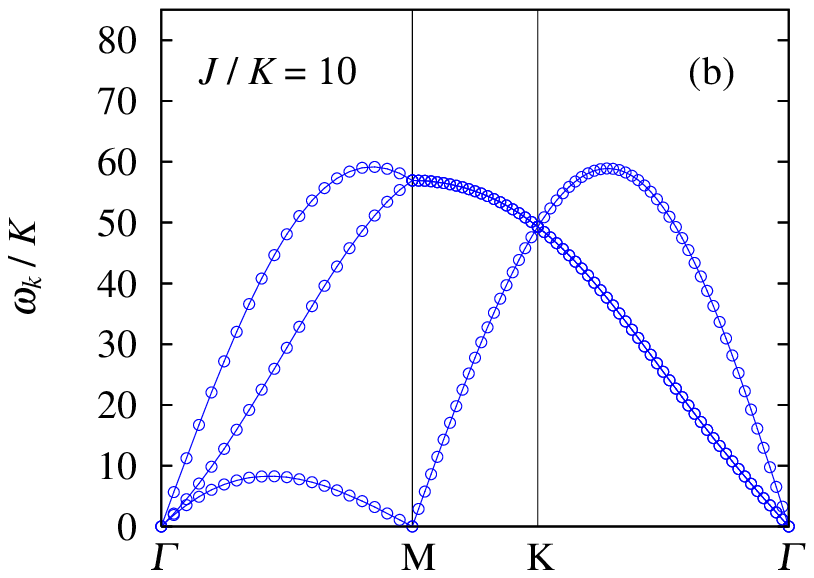}}}
 \caption{Three dispersions of the spin wave for (a) $J/K=12$ and (b)
 $J/K=10$ at $h=0$. The marks $\Gamma$, M and K denote the wave number
 vectors as shown in Fig.~\ref{zone} and the vertical axis the lines
 $\Gamma$-M-K-$\Gamma$.}
  \label{disp_h0}
\end{figure}

Next we show the result for the MSE model at $\mib{h}=0$. 
In this case the transformed Hamiltonian~(\ref{ham2}) can
be analytically diagonalized~\cite{Kubo2} as
\begin{equation}
  {\cal H}= E^{\rm cl}+\sum_{\mib{k}}(
      \omega_{\mib{k}}^{(1)}\alpha_{\mib{k}}^{\dagger}\alpha_{\mib{k}} 
    + \omega_{\mib{k}}^{(2)}\beta_{\mib{k}}^{\dagger}\beta_{\mib{k}} 
    + \omega_{\mib{k}}^{(3)}\gamma_{\mib{k}}^{\dagger}\gamma_{\mib{k}}) 
    +  E^{\rm q} \ ,
\label{ham_h=0}
\end{equation}
where the summation of the second term runs over the Brillouin zone, and
\begin{eqnarray}
 && \omega_{\mib{k}}^{(\mu)}=\sqrt{V_{\mib{k}}^{(\mu)}-W_{\mib{k}}^{(\mu)}}
   \ , \nonumber \\
 && V_{\mib{k}}^{(1)}=6(J-K)+9K\Delta_{\mib{k}}+3(J+5K) 
  \Gamma '_{\mib{k}} \ , \nonumber \\
 && W_{\mib{k}}^{(1)}=9\{ K\Delta_{\mib{k}}+(J+K) 
  \Gamma '_{\mib{k}}  \} \ ,  
\label{non-field}
\end{eqnarray}
and  $\Gamma '_{\mib{k}} ={\rm Re}(\Gamma_{\mib{k}})$, 
$V_{\mib{k}}^{(2)}=V_{\mib{k}+\mib{Q}}^{(1)}$,
$W_{\mib{k}}^{(2)}=W_{\mib{k}+\mib{Q}}^{(1)}$,
$V_{\mib{k}}^{(3)}=V_{-\mib{k}+\mib{Q}}^{(1)}$,
$W_{\mib{k}}^{(3)}=W_{-\mib{k}+\mib{Q}}^{(1)}$ 
where $\mib{Q}=(-2\pi/3, 2\pi/3)$.
The first and third terms of Eq.~(\ref{ham_h=0}) are
\begin{equation}
  E^{\rm cl}=-\frac{3}{2}(J+K)N \ , 
\end{equation}
\begin{equation}
  E^{\rm q}=\frac{1}{2}\sum_{\mu=1}^3 \sum_{\mib{k}} 
                (\omega_{\mib{k}}^{(\mu)}-V_{\mib{k}}^{(\mu)}) \ .
\label{quantum_E}
\end{equation}
We show the spin-wave spectrum for $J/K = 12$ and 10 in
Figs.~\ref{disp_h0} (a) and (b), respectively. The frequencies of
branches vanish at the $\Gamma$ point as for the AFH model. 
They have linear dispersions for small $k$ as 
\begin{eqnarray}
\omega_{\mib{k}}^{(1)} &\simeq &  3\sqrt{3(J-K)(J+2K)}\  k \ , \\
\omega_{\mib{k}}^{(2)} &\simeq& \omega_{\mib{k}}^{(3)} \\ 
\simeq && \hspace{-1cm} 3
\sqrt{6(J-K)\{(J-7K)(k_1^2 + k_2^2 ) - (J+5K) k_1 k_2 \} } \ . \nonumber
\end{eqnarray}
Figure~\ref{disp_h0} shows that the frequency of the second branch 
$\omega_{\mib{k}}^{(2)}$ vanishes at the M point, which indicates that a
second-order phase transition to the 6-sublattice structure occurs due
to the softening of the spin wave just at $J=10K$~\cite{Kubo2}. The
softening occurs at six equivalent M points in the Brillouin zone.

\begin{figure}[td]
  \centerline{\resizebox{0.4\textwidth}{!}{\includegraphics{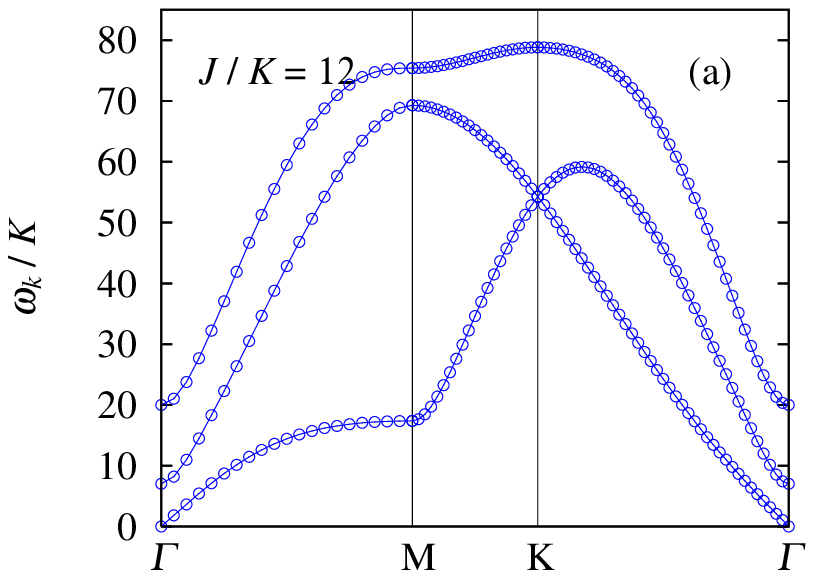}}}
  \centerline{\resizebox{0.4\textwidth}{!}{\includegraphics{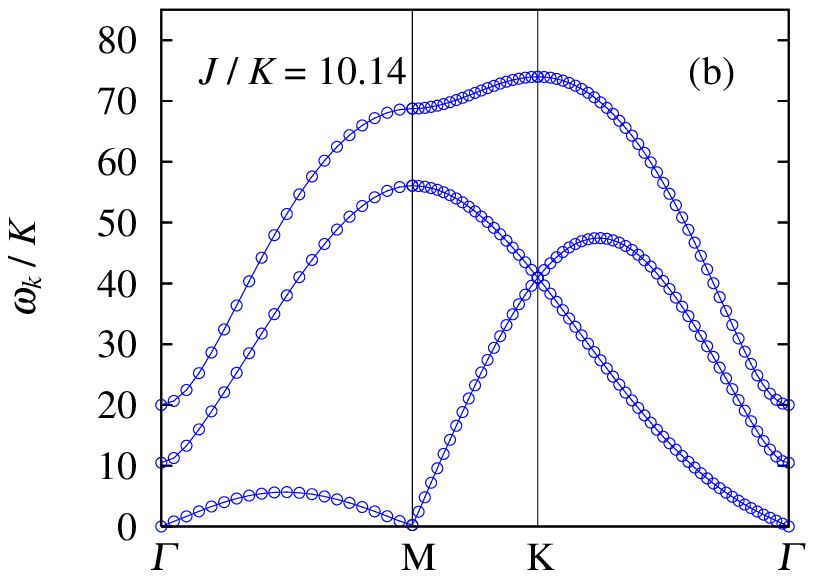}}}
  \caption{Three dispersions of the spin wave for (a) $J/K=12$ and (b)
 $J/K=10.14$ at $h/K=10$. The frequency of the lowest branch at the M
 point vanishes at $J/K \simeq 10.14$.}
  \label{disp_h10}
\end{figure}

\section{Spin-Wave Spectrum of the MSE Model in the Magnetic Field}

\begin{figure}[t]
  \centerline{\resizebox{0.5\textwidth}{!}{\includegraphics{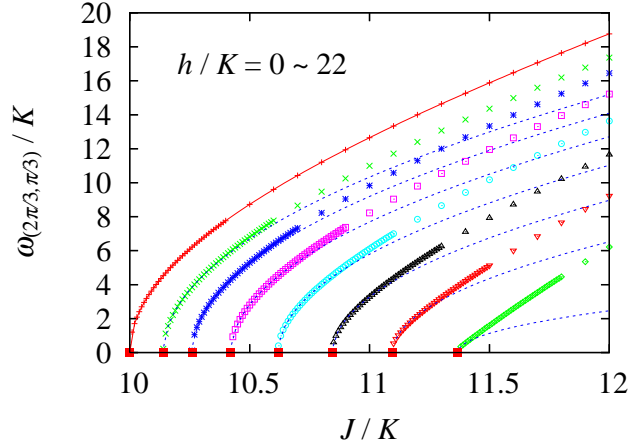}}}
  \caption{$J/K$-dependences of the frequency of the spin wave at the
 point M in the magnetic field $h/K=0$, 10, 12, 14, 16, 18, 20 and 22 from top.
 The solid marks for $\omega(2\pi/3,\pi/3)=0$ denote the phase-transition
 points. The bold line for $h/K=0$ is analytically obtained by using
 Eq.~(\ref{non-field}). The broken lines are obtained by the
 least-squares fitting with the fitting function 
 $\omega_k \propto (J-J_{\rm c})^\beta$ with $\beta=1/2$.}
  \label{soft}
\end{figure}

In Figs.~\ref{disp_h10} (a) and (b) we show the dispersion for
$J/K=12$ and 10.14 at $h=10K$. In contrast with the spectrum in the AFH
model, there is only one gapless branch at the $\Gamma$ point and the
gapless branch has a linear dispersion. The result is consistent with
the excitations from a stable ground state only with the global SO(2)
symmetry and corresponds to that the four-spin exchanges stabilize the
coplanar Y-shape state already in the mean-field approximation. The
frequency of the lowest branch decreases at the M point with decreasing
$J/K$ for a fixed $h$. It vanishes at a critical value of $J/K$ as
shown in Fig.~\ref{disp_h10} (b). The $J/K$-dependence of the frequency
of the spin wave at the point M ($\mib{k}=(2\pi/3,\pi/3)$) is shown in
Fig.~\ref{soft} for various values of $h$.
The softening of the spin wave is induced by the competition of the two-
and four-spin interactions in the magnetic field. The frequency vanishes
apparently as $\omega_k \propto (J-J_{\rm c})^\beta$. 
Here we assume the critical exponent $\beta$ to be 1/2 which results from
the general Landau theory. The softening of the spin wave leads to a
phase transition to the 6-sublattice phase in the magnetic field. 
We determine the phase boundary between the Y-shape and the 6-sublattice
phases by determining the critical value in the next section.

\section{The Phase Diagram} 

\begin{figure}[t]
  \centerline{\resizebox{0.5\textwidth}{!}{\includegraphics{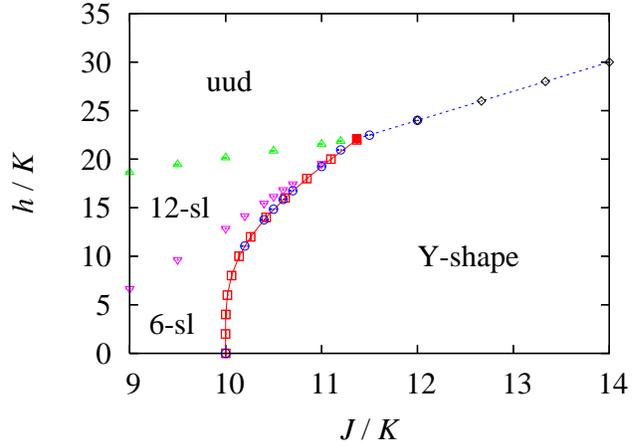}}}
  \caption{Phase diagram parametrized by the two-spin interaction and
 the magnetic field. The abbreviations of Y-shape, 6-sl, 12-sl and uud
 denote the phases with the Y-shape, 6-, 12-sublattice and uud structures. 
 The open squares are estimated by the softening of the spin wave
 at $\mib{k}=(2\pi/3,\pi/3)$. The diamonds denote the parameters at
 which the total magnetization is equal to the saturated value 1/3 of
 the uud structure. The broken line is the phase-transition line
 $h_{\rm c}=-12K+3J$. For reference, the data in ref.~13 
 estimated by the vector and scalar chiral order parameter within the
 mean-field approximation are also shown as circles and triangles. The
 solid square denotes the multi-critical point ($(J/K)_{\rm multi}$,
 $(h/K)_{\rm multi}$) $\simeq$ (11.37, 22.11). The bold line is a guide
 for the eye.}
  \label{phase}
\end{figure}

In Fig.~\ref{phase} we show the phase diagram determined from the
softening of the spin wave together with the phase boundaries 
determined by the mean-field calculations. The phase boundary between
the Y-shape and the 6-sublattice phases extends from 
$(J/K, h/K) = (10, 0)$ to the multi-critical point 
($(J/K)_{\rm multi}$, $(h/K)_{\rm multi}$) where four phase boundaries 
appear to merge.  Phase transitions from the Y-shape phase to  
the 6-sublattice and the uud phases are continuous.
On the other hand, the transition between the 6-sublattice and the
12-sublattice phases and that between the 12-sublattice and the uud
phases are discontinuous. The phase boundary between the Y-shape and the
uud phases is given by the relation (\ref{eq:hc}) and the multi-critical
point is estimated to be 
($(J/K)_{\rm multi}$, $(h/K)_{\rm multi}$) $\simeq$ (11.37, 22.11).  
Though in the AFH model the uud phase occurs only at a critical value
$h_{\rm c}=3J$ in the mean-field theory, the phase extends to a finite
region of $h$ in the MSE model. The four-spin exchange stabilizes the
uud state. It was shown that the thermal or quantum fluctuations
stabilize the uud state~\cite{Kawamura,Chubukov3}. The four-spin
exchange has a same effect and as a result magnifies the 1/3
magnetization plateau. 

In the phase diagram, we also mark the phase transition points among the
Y-shape, 6-, 12-sublattice and uud phases, which are estimated by the
vector and scalar chiral order parameters within the mean-field
approximation.  

\begin{figure}[t]
  \centerline{\resizebox{0.4\textwidth}{!}{\includegraphics{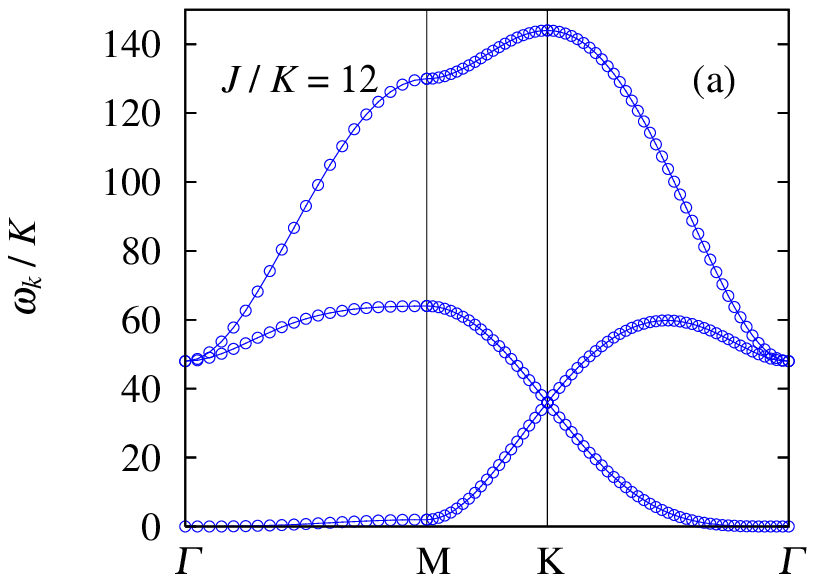}}}
  \centerline{\resizebox{0.4\textwidth}{!}{\includegraphics{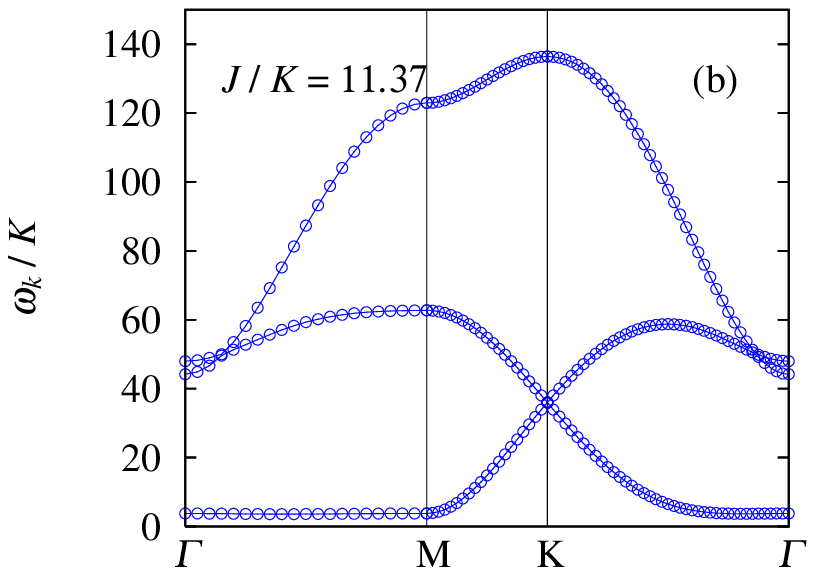}}}
  \caption{Three dispersions of the spin wave for (a) $J/K=12$ and
(b) $J/K=11.37$ ($\simeq (J/K)_{\rm multi}$) at $h/K=24$. The flat
 branch is observed along the $\Gamma$-M line.}
 \label{disp_uud_h}
\end{figure}

\begin{figure}[t]
  \centerline{\resizebox{0.5\textwidth}{!}{\includegraphics{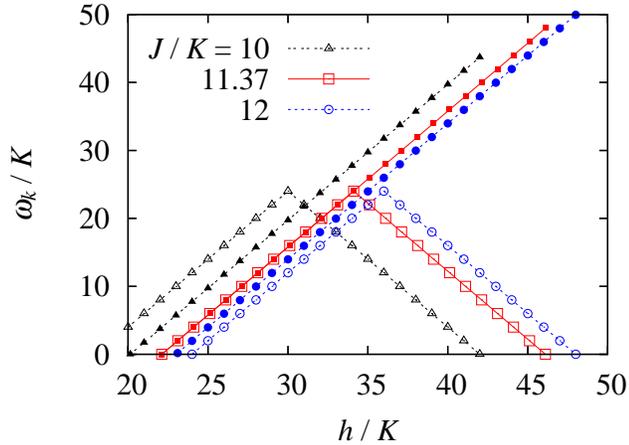}}}
  \caption{Magnetic-field dependences of the frequencies of the lowest
 branch at $\Gamma$ (open) and M (solid) points for $J/K=10$, 11.37
 ($\simeq (J/K)_{\rm multi}$) and 12. All the lines are guides for the eye.}
  \label{gap}
\end{figure}

Peculiar properties are expected at the multi-critical point. Since the
spin-wave analysis is broken up just at the multi-critical point, we
investigate systems close to the point. In Fig.~\ref{disp_uud_h} (a)
we show the dispersion of the spin wave for $J=12K$ and $h=24K$ just on the
phase-transition line between the Y-shape and the uud phases, which is
calculated by the spin-wave theory assuming the uud state as the ground
state. We found that the lowest gapless branch has a tendency to be flat
along the $\Gamma$-M line. Actually, for 
$J/K=11.37 \simeq (J/K)_{\rm multi}$ the branch along the $\Gamma$-M
line is flat as shown in Fig.~\ref{disp_uud_h} (b). As the magnetic
field is increased from the value on the phase-transition line, the
frequency of the flat lowest branch for 
$J/K=11.37 \simeq (J/K)_{\rm multi}$ linearly increases with $h/K$ in
the range $22.11 < h/K \nle 34$ as demonstrated in Fig.~\ref{gap}. 
Since the second branch at the $\Gamma$ point as shown in
Fig.~\ref{disp_uud_h} (b) crosses the lowest branch at $h/K \simeq 34$,
the $h$-dependence of the frequency of the lowest branch at the $\Gamma$
point has an inflection at $h/K \simeq 34$. 
When the value of $J/K$ is shifted from the multi-critical point, the
flatness due to the balance of the two- and four-spin exchanges is
broken up, though the linearity still holds. In the uud phase
the frequency of the lowest branch has the relation $\omega_k=2(h-h_1)$
and $2(h_2-h)$ for $h_1 < h < (h_1+h_2)/2$ and $(h_1+h_2)/2 < h < h_2$,
respectively, where $h_1$ and $h_2$ are the magnetic fields of the lower
and the upper phase boundaries of the uud phase. 
The linearity of the $h$-dependence of the frequency has been also
obtained in the uud phase for the quantum AFH model~\cite{Chubukov3}.

\section{Ground-State Energy and Sublattice Magnetization}

\begin{figure}[t]
  \centerline{\resizebox{0.5\textwidth}{!}{\includegraphics{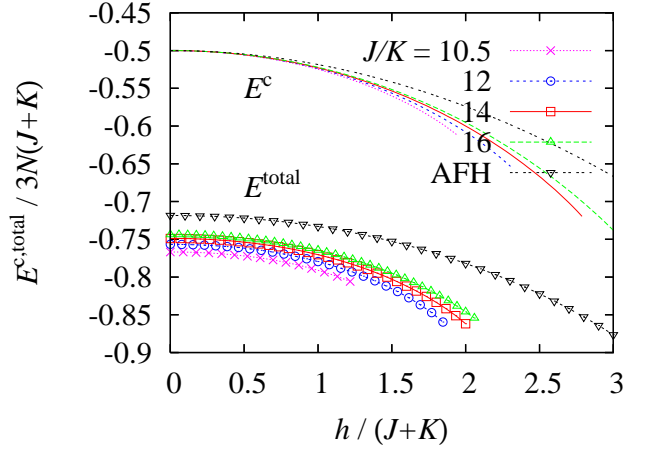}}}
  \caption{Magnetic-field dependences of the classical part of the
 energy $E^{\rm c}$ and the total energy $E^{\rm total}$ per bond
 normalized by $J+K$. For comparison, the data of the AFH model ($K=0$)
 are also shown. The lines on the marks are guides for the eye.}  
\label{enec}
\end{figure}

\begin{figure}[t]
  \centerline{\resizebox{0.5\textwidth}{!}{\includegraphics{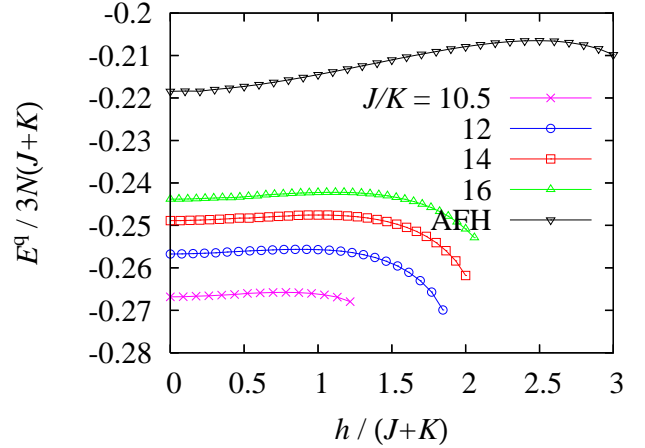}}}
  \caption{Magnetic-field dependences of the quantum correction of the
 energy per bond normalized by $J+K$. The lines are guides for the eye.}
\label{eneq}
\end{figure}

The total energy of the ground state is evaluated by 
$E^{\rm c}+E^{\rm q}$, where $E^{\rm c}$ is the classical energy of
Eq.~(\ref{classical_E}) and can be analytically calculated, 
and $E^{\rm q}$ is the quantum energy corresponding to, e.g.,
Eq.~(\ref{quantum_E}) for $h=0$. The magnetic-field dependences of
$E^{\rm c}$ and $E^{\rm q}$ per bonds are shown in Figs.~(\ref{enec})
and (\ref{eneq}), respectively, where we normalize them by using $J+K$
to compare with the values of the AFH model. It is found that, as $J/K$
is decreased, the quantum correction of the energy is increased.

The total magnetization is defined by using the expectation value of the
spin $z$ component, i.e.,
\begin{equation}
  M=-\frac{1}{N}\sum_{{\rm all}~i} \langle \sigma_i^z \rangle \ .
\label{totalm}
\end{equation}
Note that we take the direction of the magnetic field anti-parallel to
the $z$ direction. The sublattice magnetizations of the A and B
sublattices are defined by 
\begin{equation}
  m_{\rm A}=-\frac{3}{N}\sum_{i \in A} \langle \sigma_i^z \rangle
           = -1+\Delta m_{\rm A} \ ,
\label{ma}
\end{equation}
\begin{equation}
  m_{\rm B}=-\frac{3}{N}\sum_{j \in B} \langle \sigma_j^z \rangle
           = \beta (1-\Delta m_{\rm B}) \ ,
\label{mb}
\end{equation}
respectively, where from Eqs.~(\ref{HP_A}) and (\ref{HP_B}),
\begin{equation}
  \Delta m_{\rm A} = \frac{6}{N}\sum_{i \in A} \langle
                              a_i^{\dagger} a_i \rangle  \ ,
\end{equation}
\begin{equation}
  \Delta m_{\rm B} = \frac{6}{N}\sum_{j \in B} \langle
                              b_j^{\dagger} b_j \rangle  \ .
\end{equation}
From a symmetry of the present system, the sublattice magnetization of
the C sublattice is same as that of the B sublattice.
The total magnetization is written from Eqs.~(\ref{totalm})-(\ref{mb}),
i.e.,
\begin{equation}
   M = M^{\rm cl} + \Delta M \ ,
\end{equation}
where
\begin{equation}
   M^{\rm cl} = -\frac{1}{3} (1-2\beta) \ ,
\end{equation}
\begin{equation}
  \Delta M = \frac{1}{3} (\Delta m_{\rm A} - 2\beta \Delta
   m_{\rm B}) \ .
\end{equation}

\begin{figure}[t]
  \centerline{\resizebox{0.5\textwidth}{!}{\includegraphics{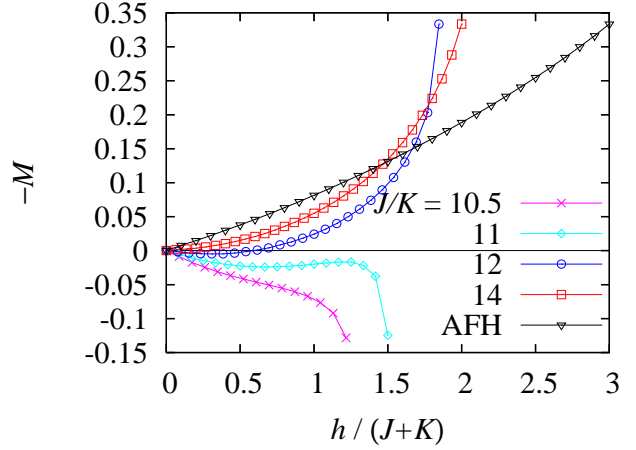}}}
  \caption{Dependences of the total magnetization on the magnetic field
 normalized by $J+K$. The data described as AFH are results for $K=0$.}
\label{total_mag}
\end{figure}

\begin{figure}[t]
  \centerline{\resizebox{0.5\textwidth}{!}{\includegraphics{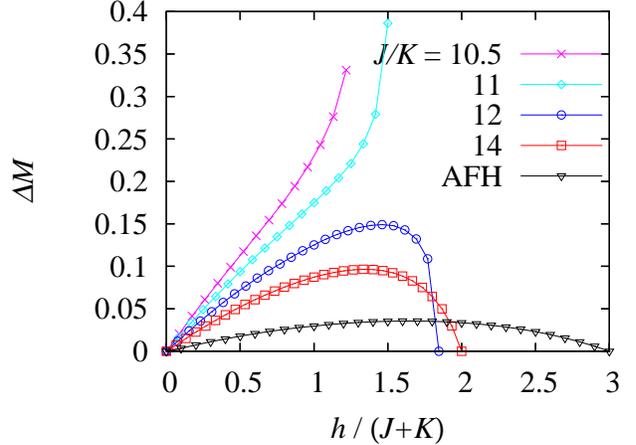}}}
  \caption{Dependences of the quantum correction of the total
 magnetization on the magnetic field normalized by $J+K$.}
\label{total_qcorr}
\end{figure}

The magnetic-field dependences of $M$ for various values of $J/K$ are
shown in Fig~\ref{total_mag}. When $h=0$, the total magnetization is to
be zero since the ground state has the 120$^\circ$ structure for all
systems. For the AFH system, the value of $-M$ is increased with $h$ and
saturates to $-M=1/3$ at $h/J=3$, where the phase transition to the uud
phase occurs. While we can obtain the same behavior as the AFH system for
$J/K=14$, the values of $-M$ become negative for small $h/(J+K)$ for
$10.5 \le J/K \le 12$. This result shows that the spin-wave theory on
the sublattice magnetization is broken up due to strong quantum
fluctuations in the region. 
We also show the $h$-dependence of the quantum correction
$\Delta M$ in Fig.~\ref{total_qcorr}. The quantum corrections vanish
both at $h=0$ and the phase transition point to the uud phase. 

\begin{figure}[t]
  \centerline{\resizebox{0.5\textwidth}{!}{\includegraphics{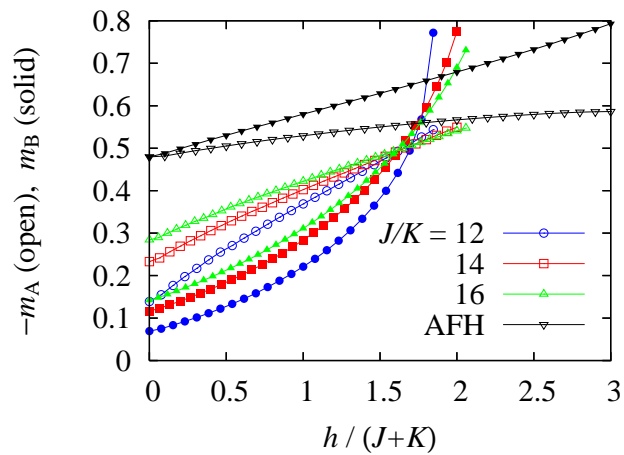}}}
  \caption{Dependences of the sublattice magnetizations on the magnetic
 field normalized by $J+K$. The open and solid marks denote $-m_{\rm A}$
 and $m_{\rm B}$, respectively.}
\label{sub_mag}
\end{figure}

\begin{figure}[t]
  \centerline{\resizebox{0.5\textwidth}{!}{\includegraphics{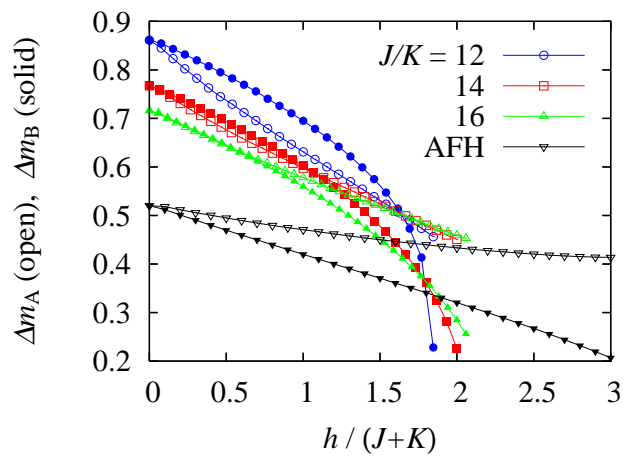}}}
  \caption{Dependences of the quantum corrections of the sublattice
 magnetizations on the magnetic field. The open and solid marks denote
 $\Delta m_{\rm A}$ and $\Delta m_{\rm B}$, respectively.}
\label{qcorr}
\end{figure}

The magnetic-field dependences of the sublattice magnetizations
of the A and B sublattices for various values of $J/K$ are shown in
Fig.~\ref{sub_mag}. The values of $-m_{\rm A}$ and $m_{\rm B}$ are
decreased with $J/K$ and $h/(J+K)$ except near the phase transition
point to the uud phase. The reduction shows that the quantum effects are
enhanced by the addition of the four-spin interaction in the MSE
model. We also show the dependence of the quantum corrections of the
sublattice magnetizations on the magnetic field in Fig.~\ref{qcorr}. The
quantum corrections are larger for smaller $J/K$ and larger $h/(J+K)$.
We can obtain an interesting tendency of the size relation between
$\Delta m_{\rm A}$ and $\Delta m_{\rm B}$ within the present spin-wave
approximation. In the AFH system with $K=0$, the value of 
$\Delta m_{\rm A}$ is larger than that of $\Delta m_{\rm B}$ except for
$h=0$. The result suggests that the sublattice magnetization along the
axis anti-parallel to the direction of the magnetic field is most
sensitive to quantum fluctuation. In the MSE system with the large
four-spin interaction and small magnetic field, on the other hand, 
the sublattice magnetization whose quantization axis tends to parallel
to the direction of the magnetic field is most sensitive to quantum
fluctuation.

\section{Summary and Discussions}

In the present paper, we studied quantum effects on the 3-sublattice
structures in the $S=1/2$ multiple-spin exchange model with two-, three-
and four-spin exchange interactions on the triangular lattice in the
magnetic field. By using the linear spin-wave theory, we found that the
coplanar Y-shape state is stable as the ground state of the quantum
system, though the four-spin interaction leads to the instability as the
softening of the spin wave and the phase transition to the 6-sublattice
phase occurs.

We had arguments in the framework of the linear spin-wave theory. Within
the approximation, the magnetization becomes negative in the parameter
region where the four-spin exchange interaction is dominant. We should
take into account higher-order corrections of the spin wave to
estimate the magnetization in the region. 

The exact diagonalization study of finite clusters predicted that the
ground state is a spin-liquid state with a spin gap filled with a large
number of singlet states for parameters corresponding to our
6-sublattice phase~\cite{LiMing}. The character of the ground state in
this parameter region is, however, still not clarified. In order to
numerically examine the possibility of the 6-sublattice phase, we would
need the exact diagonalization of clusters with larger sizes. The
spin-wave analysis assuming the 6-sublattice structure as the ground
state is another interesting problem remained for the future work.

\acknowledgements

We acknowledge useful discussions with Y. Uchihira.
This work is partly supported by Grants-in-Aid for Scientific Research
Programs (No.~17071011, No.~17540339, No.~18740239)
and 21st Century COE program from the Ministry of
Education, Culture, Sports, Science and Technology of Japan.


\begin{thebibliography}{99}
\bibitem{y-shape} It was called umbrella state in ref. 5.
\end{thebibliography}
\end{document}